\documentclass[
article,
superscriptaddress,
preprint,
pra,
aps
]{revtex4-2}

\usepackage[colorlinks=true, linkcolor=blue, urlcolor=blue, citecolor=blue, anchorcolor=blue]{hyperref}

\usepackage[utf8]{inputenc}
\usepackage{amsmath}
\usepackage{graphicx}
\usepackage{bm}
\usepackage{float}

\begin{document}

\title{Universal structure of propagation-invariant optical pulses}

\author{R. Almeida}
\email{rafaelrussoalmeida@tecnico.ulisboa.pt}
\affiliation{GoLP/Instituto de Plasmas e Fusão Nuclear, Instituto Superior Técnico, Universidade de Lisboa, Lisbon, Portugal}
\author{D. Ramsey}
\affiliation{University of Rochester, Laboratory for Laser Energetics, Rochester, New York, 14623 USA}
\author{A.F.  Abouraddy}
\affiliation{CREOL, The College of Optics \& Photonics, University of Central Florida, Orlando, FL 32816, USA}
\author{J.P. Palastro}
\affiliation{University of Rochester, Laboratory for Laser Energetics, Rochester, New York, 14623 USA}
\author{J. Vieira}
\email{jorge.vieira@tecnico.ulisboa.pt}
\affiliation{GoLP/Instituto de Plasmas e Fusão Nuclear, Instituto Superior Técnico, Universidade de Lisboa, Lisbon, Portugal}

\date{\today}

\begin{abstract}
Space-time structuring of light—where spatial and temporal degrees of freedom are deliberately coupled and controlled—is an emerging area of optics that enables novel configurations of electromagnetic fields. Of particular importance for applications are optical pulses whose peak intensity travels at an arbitrary, tunable velocity while maintaining its spatiotemporal profile. Space-time wave packets and the ideal flying focus are two prominent realizations of these pulses. Here, we show that these realizations share an identical spatiotemporal field structure, and that this structure represents a universal solution for constant-velocity, propagation-invariant pulses.
\end{abstract}

\maketitle

The development of techniques to simultaneously manipulate and couple the spatial and temporal degrees of freedom of optical pulses \cite{shen_roadmap_2023} has enabled diverse and novel structures, including spatiotemporal optical vortices \cite{jhajj_spatiotemporal_2016, hancock_free-space_2019, chong_generation_2020, bliokh_spatiotemporal_2021}, light springs \cite{pariente_spatio-temporal_2015, piccardo_broadband_2023}, and toroidal pulses \cite{zdagkas_observation_2022, wan_toroidal_2022}. Two particular examples---space-time wave packets (STWPs) \cite{kondakci_diffraction-free_2017} and the flying focus (FF) \cite{froula_spatiotemporal_2018}---feature a tunable and arbitrary velocity intensity peak, providing a powerful new tool for light-matter interactions. The intensity peak of a STWP is propagation invariant in both free-space \cite{bhaduri_broadband_2019, yessenov_space-time_2022} and dispersive media \cite{hall_canceling_2023, he_nondispersive_2022}, the latter having been used for the realization of optical delay lines \cite{yessenov_free-space_2020}. The  FF promises to overcome dephasing in laser wakefield accelerators \cite{caizergues_phase-locked_2020, palastro_dephasingless_2020}, increase the rate of frequency up-shifting in photon acceleration \cite{howard_photon_2019,franke_optical_2021}, and facilitate the formation of long plasma channels by mitigating ionization refraction \cite{palastro_ionization_2018,Turnbull2018}. 

Both STWPs \cite{kondakci_diffraction-free_2017} and the FF \cite{froula_spatiotemporal_2018} use spatiotemporal couplings to control the velocity of the peak intensity. The underlying concepts and methods, however, are distinct. With STWPs, each spatial wavenumber is assigned to a single temporal frequency. With the FF, the time and location where different parts of a pulse focus are dynamically controlled to create a moving focal point. Unlike STWPs, the FF does not generally produce a propagation-invariant intensity peak. However, a recently discovered variant of the FF, referred to as the ideal flying focus (IFF), does produce such a peak \cite{franke_optical_2021, simpson_spatiotemporal_2022, ramsey_exact_2023}. This raises the question: are STWPs and the IFF truly unique realizations of space-time structured light?

In this Letter, we show that STWPs and the IFF share an identical spatiotemporal field structure despite being conceptually distinct and synthesized through different methods. In fact, this structure is universal and applies to any constant-velocity, propagation-invariant pulse. The universality is demonstrated by applying Lorentz boosts to the spectra of simpler field structures. An extension of this analysis shows that finite-energy pulses can exhibit a propagation-invariant intensity peak over a finite distance. Thus, the propagation invariance of STWPs and the IFF does not require idealized, infinite-energy pulses.

\begin{figure}[t!]
    \centering
    \includegraphics[width=8.8cm]{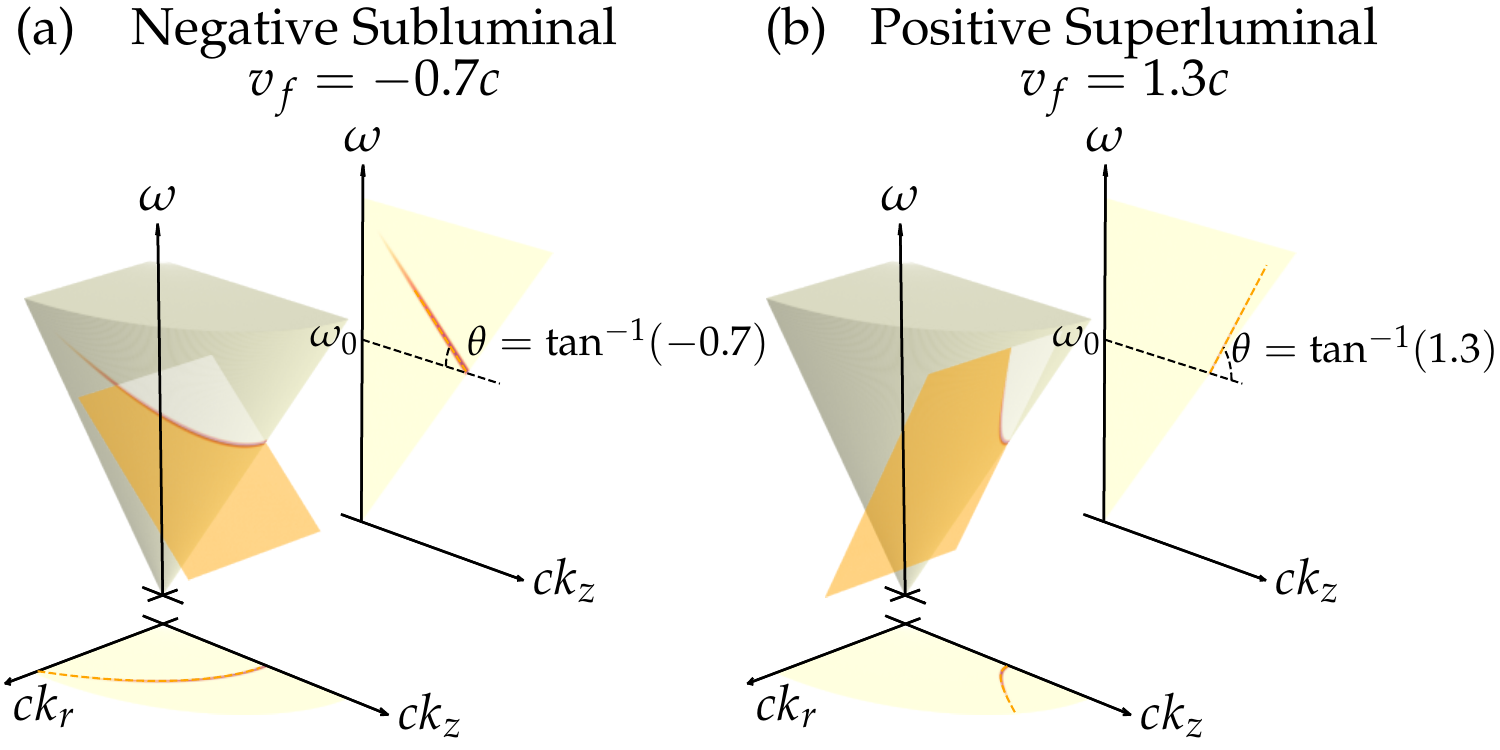}
    \caption{A wave packet composed of frequencies and wavenumbers lying at the intersection of the light cone and a tilted plane will have a group velocity $v_f$ determined by the tilt angle of the plane $\theta$: (a) $v_f=-0.7c$ and (b) $v_f=1.3c$. The $\omega-ck_z$ projections confirm that $\partial\omega/\partial k_z=v_f$, while the $ck_r-ck_z$ projections show the spectral content for each velocity.}
    \label{fig:stwp_cones}
\end{figure}

\textbf{\textit{Space-time Wave Packets}} - STWPs \cite{kondakci_optical_2019, yessenov_space-time_2022} are a class of space-time structured pulses with a well-defined correlation between their wave vectors and temporal frequencies. By tuning this correlation, the group velocity of a STWP can take any value, from positive to negative superluminal \cite{bhaduri_broadband_2019}. The spectral representation of a STWP can be visualized on the light-cone---a 4D hypersurface in $(\omega, \mathbf{k})$ space defined by the vacuum dispersion relation $\omega=c|\mathbf{k}|$, where $c$ is the vacuum speed of light, $\omega$ the frequency, and $\mathbf{k}=(k_x,k_y,k_z)$ the wave vector in Cartesian space $(x,y,z)$. For cylindrically symmetric STWPs, a single transverse wavenumber $k_r\equiv(k_x^2+k_y^2)^{1/2}$ is sufficient, allowing for a 3D representation of the light cone in $(ck_r,ck_z,\omega)$ (Fig.~\ref{fig:stwp_cones}). The spectra of all realizable optical fields reside on the surface of the light-cone. An arbitrary group velocity $v_f=\partial\omega/\partial k_z$ can be achieved by choosing only those frequencies and longitudinal wavenumbers that satisfy $\omega=\omega_{\mathrm{0}}+(k_z-\omega_{\mathrm{0}}/c)v_f$, where $\omega_{\mathrm{0}}$ is a fixed frequency. This defines a plane that is tilted at an angle $\theta=\arctan(v_f/c)$ with respect to the $k_z$ axis. The intersection of this plane with the light cone is a conic section. Figure~\ref{fig:stwp_cones} shows that the intersection is equivalent to having a longitudinal wavenumber $k_z$ that depends on, i.e., is correlated with, the transverse wavenumber $k_r$. This correlation can also be derived by applying spectral Lorentz boosts to generic beams \cite{belanger_lorentz_1986,longhi_gaussian_2004,yessenov_relativistic_2023,yessenov_experimental_2024}. 

\textbf{\textit{The Ideal Flying Focus}} - FF techniques \cite{froula_spatiotemporal_2018, Pigeon2024} tailor the trajectory of a focus by controlling the focal time and location of each frequency, temporal slice, or annulus of a pulse. The original FF used chromatic focusing of a chirped laser pulse. At every location along the extended focal region of the chromatic lens, the moving intensity peak comprises a different portion of the overall spectrum. As a result, the intensity peak produced by the original ``chromatic'' FF is only approximately propagation invariant. The more recently proposed IFF remedies this by substituting the chirped pulse and chromatic lens for an unchirped pulse and a lens with a time-dependent focal length \cite{simpson_spatiotemporal_2022}. The exact electromagnetic fields of an IFF pulse \cite{ramsey_exact_2023} were derived by superposing multipole spherical or hyperbolic wave solutions, applying the complex source-point method \cite{sheppard_electromagnetic_1999, heyman_gaussian_2001}, and Lorentz boosting the result. The fields obtained using this method were shown to be exactly those of a pulse focused by a lens with a time-dependent focal length \cite{ramsey_exact_2023}. 

\begin{figure}[h]
    \centering
    \includegraphics[width=8.8cm]{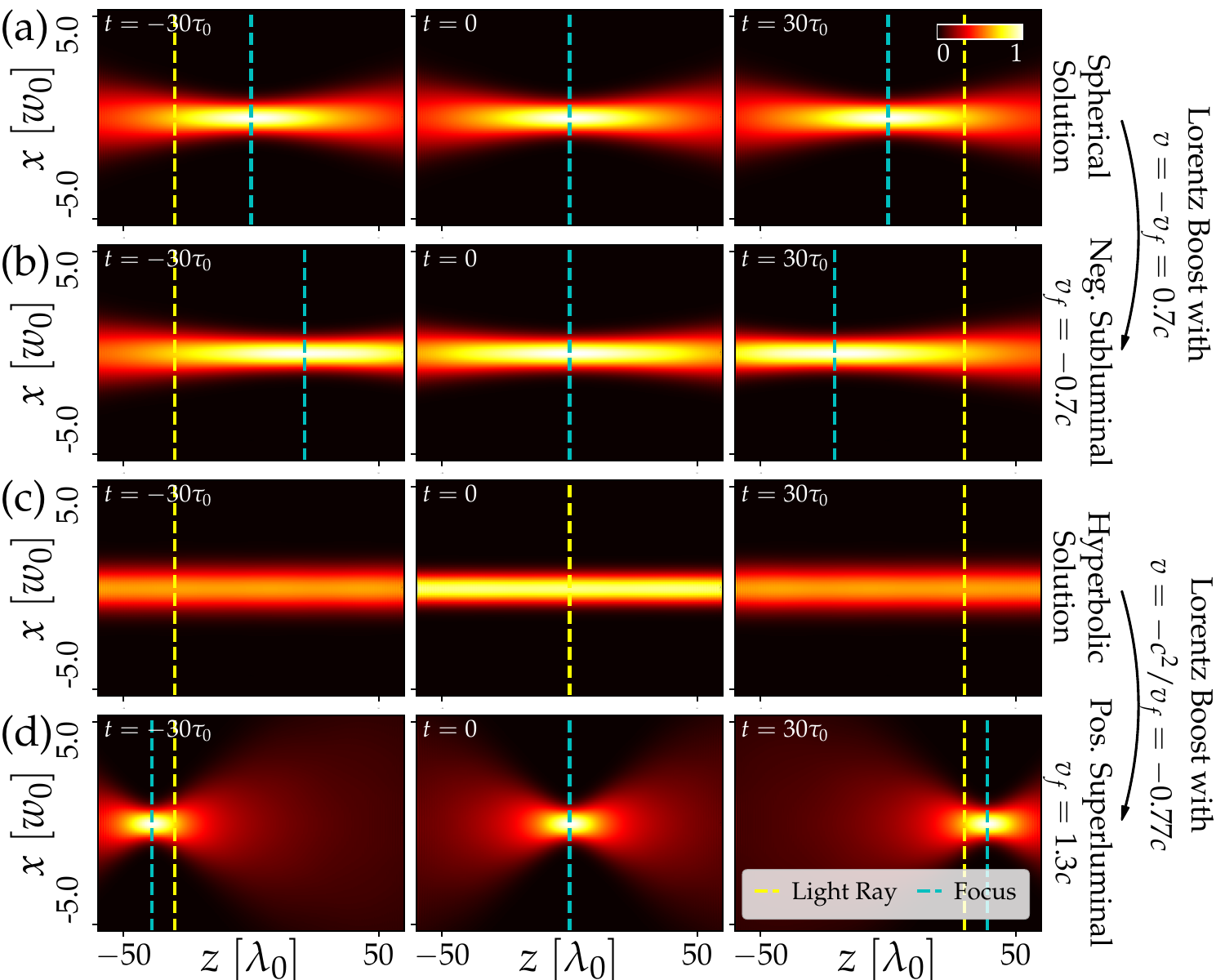}
    \caption{Lorentz boosts applied to known fields produces constant-velocity, propagation invariant pulses. (a) A gaussian wave packet with a stationary focus becomes a (b) propagation-invariant wave packet traveling at $v_f = -0.7c$ after a Lorentz boost with $\boldsymbol{v}=0.7c\mathbf{\hat z}$. (c) A wave packet that focus everywhere in space at the same time becomes a (d) propagation-invariant wave packet traveling at $v_f =1.3c$ after a Lorentz boost with $\boldsymbol{v}=-0.77c\mathbf{\hat z}$. In these examples, the spot size $w_0=3 c\tau_0$, where $\tau_0$ is the laser period.}
    \label{fig:exact_boost}
\end{figure}

\textbf{\textit{Equivalence of IFF and STWPs}} - The connection of STWPs and the IFF to Lorentz boosts of known solutions to Maxwell's equations motivates a search for a more-general underlying structure. Subluminal ($\smash{|v_f| < c}$) and superluminal ($\smash{|v_f| > c}$) propagation-invariant wave packets can be constructed by Lorentz boosting a known optical field \cite{belanger_lorentz_1986, longhi_gaussian_2004}. Consider Fig.~\ref{fig:exact_boost}(a) which shows a standard solution of Maxwell's equations for a wave packet with a stationary focus (i.e., a spherical wave solution). A Lorentz boost to a frame moving at $v = - v_f$ in the $z$ direction produces a focus that moves subluminally at $v_f$ in the new frame [Fig.~\ref{fig:exact_boost}(b)]. This can be extended to superluminal foci by starting with a solution where the focus occurs at the same time at every point in space (i.e., a hyperbolic wave solution) [Fig.~\ref{fig:exact_boost}(c)]. Boosting this solution to a frame moving at $v = -c^2/v_f$ produces a focus that moves superluminally at $v_f$ in the new frame. [Fig.~\ref{fig:exact_boost}(d)].

The equivalence of STWPs and the IFF will be demonstrated by applying these Lorentz boosts in spectral space. To begin, note that the phase of a light wave is a Lorentz invariant quantity given by the dot product of the frequency and position four-vectors: $k^\mu x_\mu$. As a result, a boost can be enacted on either the real-space coordinates $x^\mu$ or on the reciprocal $k^\mu$. The inertial frames of interest are the ``static'' frame of the focus denoted by $S$ where $\smash{ck^{S,\mu}=(\omega^S, ck_z^S, ck_x^S, ck_y^S)}$ and a ``laboratory'' frame where $\smash{ck^{\mu}=(\omega, ck_z, ck_x, ck_y)}$. 

\begin{figure}[t!]
    \centering
    \includegraphics[width=8.8cm]{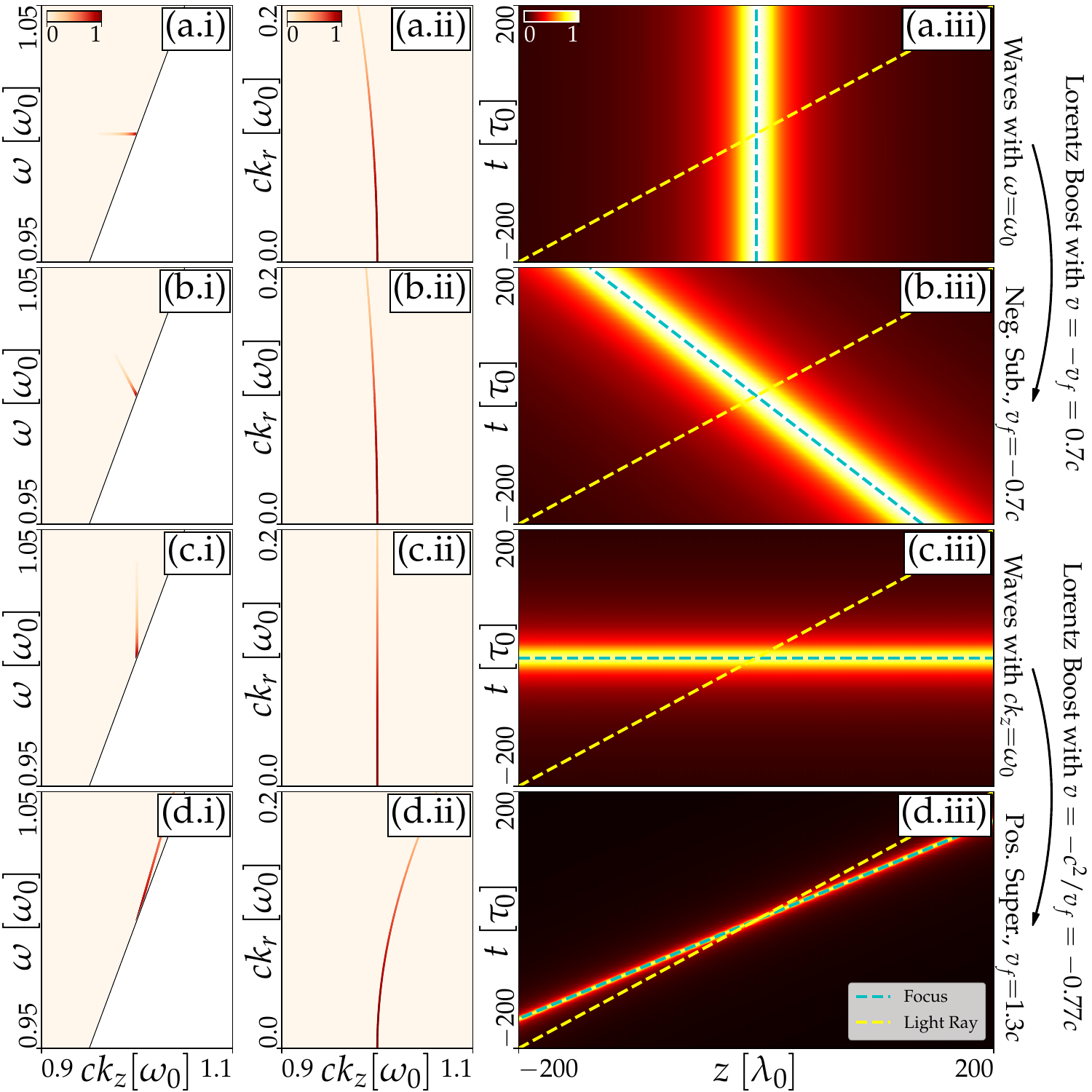}
    \caption{(i,ii) The spectral content and (iii) intensity of wave packets in the (a,c) static and (b,d) laboratory frames at $x=y=0$. (a) A wave packet that focuses at the same location for all time is Lorentz boosted to (b) a subluminal wave packet with $v_f=-0.7c$. (c) A wave packet that focuses at the same time everywhere in space is Lorentz boosted to (d) a superluminal wave packet with $v_f=1.3c$. In these examples, the wave packets have a spot size $w_0=3 c\tau_0$, where $\tau_0$ is the laser period.}
    \label{fig:flyingfocus}
\end{figure}

A subluminal ($|v_f| < c$) focus in the laboratory frame can be generated by boosting the fields of a monochromatic wave packet that focuses at the same location for all times in the static frame $S$. The spectrum of a monochromatic wave packet is restricted to a circle of constant radius $\omega_{\mathrm{0}}^S$, such that $\omega^S = \omega_{\mathrm{0}}^S$, as seen in Fig.~\ref{fig:flyingfocus}(a). If the static frame is moving at a velocity $v = -v_f$ with respect to the laboratory frame, the focus will move at a velocity $v_f$ in the laboratory frame. The frequency four-vector in the laboratory frame is then
\begin{equation}\label{eq:4freq-time-Lab}
     ck^\mu = \begin{pmatrix}
        \gamma( \omega_{0}^S {+} v_fk_z^S ), &
        \gamma( ck_z^S {+} \beta_f \omega_{0}^S ), & 
        ck_x^S, & ck_y^S 
    \end{pmatrix},
\end{equation}
where $\beta_f \equiv v_f/c$ and $\smash{\gamma\equiv(1-\beta_f^2)^{-1/2}}$. A main frequency $\omega_0$ in the laboratory frame is obtained by setting $\smash{\omega_{0}^S=\gamma(1-\beta_f)\omega_{0}}$. Using this expression and the inverse Lorentz boost $ck_z^S=\gamma(ck_z-\beta_f\omega)$ in $\smash{ck^0 = \omega = \gamma( \omega_{0}^S + v_fk_z^S )}$ yields
\begin{equation}\label{eq:subluminal-arg}
     \omega = \omega_{0} + (k_z - \omega_0/c)v_f,\quad |v_f|<c. 
\end{equation}
Hence, a wave packet whose focus travels at a subluminal velocity is described by a superposition of plane waves with frequencies defined by \eqref{eq:subluminal-arg}, as seen in Fig.~\ref{fig:flyingfocus}(b). Explicit calculation of the group velocity verifies that it is indeed $v_f$: 
\begin{equation}\label{eq:explicit-velocity-subluminal}
    c\frac{\partial ck^0}{\partial ck^1} = c\frac{\partial [ \gamma( \omega_{0}^S {+} v_fk_z^S ) ]}{\partial [\gamma(ck_z^S {+} \beta_f \omega_{0}^S )]} = v_f.
\end{equation}

A superluminal ($|v_f| > c$) focus in the laboratory frame can be generated by boosting the fields of a wave packet that focuses everywhere in space at the same time in the static frame $S$. The spectrum of such a wave packet is centered on a constant longitudinal wavenumber $\omega_{0}^S/c$, such that $ck_z^S = \omega_{0}^S$ as seen in Fig.~\ref{fig:flyingfocus}(c). If the static frame is moving at a velocity $v = -c/\beta_f$ with respect to the laboratory frame, the focus will move at $v_f$ in the laboratory frame. In this case, the frequency four-vector in the laboratory frame is 
\begin{equation}\label{eq:4freq-space-Lab}
    ck^\mu {=} \begin{pmatrix}
        \gamma( \omega^S {+} \beta^{-1}_f \omega_{0}^S ), & 
        \gamma( \omega_{0}^S {+} \beta^{-1}_f \omega^S ), & 
        ck_x^S, & ck_y^S
    \end{pmatrix},
\end{equation}
with $\smash{\gamma\equiv(1-\beta_f^{-2})^{-1/2}}$. Employing the same reasoning as in the subluminal case, but with $\smash{\omega_{0}^S=\gamma(1-\beta_f^{-1})\omega_{0}}$ and $\smash{\omega^S=\gamma(\omega-\beta^{-1}_f ck_z)}$, yields
\begin{equation}\label{eq:superluminal-arg}
\omega = \omega_{0} + (k_z - \omega_0/c)v_f,\quad |v_f|>c. 
\end{equation}
Hence, a wave packet whose focus travels at a superluminal velocity is also a superposition of plane waves with frequencies defined by \eqref{eq:superluminal-arg}, as seen in Fig.~\ref{fig:flyingfocus}(d). As in the subluminal case, explicit calculation of the group velocity produces the expected result $v_f$ [\eqref{eq:explicit-velocity-subluminal}].

The expressions for $\omega$ in the subluminal and superluminal cases are identical [\eqref{eq:subluminal-arg} and \eqref{eq:superluminal-arg}]. These expressions are also identical to the condition on the frequencies (i.e., the plane in Fig.~\ref{fig:stwp_cones}) required for STWPs discussed earlier. Thus, the IFF and STWPs have an equivalent spectral structure defined by the intersection of the light cone and a plane. Note that this structure does not restrict the longitudinal profile of a moving focus to the Lorentzian profile produced by a focused Gaussian beam. The longitudinal profile can be shaped by superposing transverse beam modes. While spectral Lorentz boosts have been applied to more general beams in the context of STWPs \cite{yessenov_relativistic_2023, yessenov_experimental_2024}, the connection to a moving focus was not made.

\textbf{\textit{Universal structure of propagation invariant wave packets}} - The equivalence of the IFF and STWPs extends to all wave packets that feature a constant-velocity, propagation-invariant intensity peak. Consider the transverse electric field $E(x,y,z,t)$ of a propagation-invariant wave packet traveling at $v_f$ in the $z$ direction. Over a time $t_0$, the wave packet travels a distance $v_f t_0$. Up to a global phase shift, the wave packet maintains its structure, such that
\begin{equation}
    E(x,y,z,t)=E(x,y,z-v_ft_{0},t+t_{0})\ e^{i\Omega_{0}t_0}.
\end{equation}
Taking the Fourier transform of this equality and applying the Fourier-shifting property yields
\begin{equation} \label{eq:ForEq}
    \widetilde E(\mathbf{k},\omega) =
    \widetilde E(\mathbf{k},\omega)\ e^{i(k_z v_f-\omega +\Omega_{0} )t_{0}}.
\end{equation}
For non-zero $\widetilde E(\mathbf{k},\omega)$ and $t_0$, the equality in \eqref{eq:ForEq} can only be satisfied if $\omega=k_z v_f +\Omega_{0}$. Because $\Omega_{0}t_{0}$ is an arbitrary global phase shift, $\Omega_{0}$ can be freely chosen. Setting $\Omega_{0}=\omega_{0}(1-\beta_f)$ recovers the original condition for STWPs, \eqref{eq:subluminal-arg}, and \eqref{eq:superluminal-arg}: $\omega = \omega_{0} + (k_z - \omega_0/c)v_f$. Thus, for a wave packet to be shape invariant, it must take the form:
\begin{equation}\label{eq:infinite-general}
    E(\mathbf{x}, t) {=} {\int} f(\mathbf{k},\omega)\ \delta[\Omega(\omega,k_z)]\delta(\omega-c|\mathbf{k}|)\ e^{i(\mathbf{k}\cdot\mathbf{x}-\omega t)}d\mathbf{k}d\omega,
\end{equation} 
where $f(\mathbf{k},\omega)$ can be any function that determines the shape of the wave packet and the $\delta$-functions ensure $\Omega(\omega,k_z) \equiv \omega - \omega_{0} - (k_z - \omega_0/c)v_f = 0 $ and $\omega = c|\mathbf{k}|$. 

\begin{figure}[t!]
    \centering
    \includegraphics[width=8.8cm]{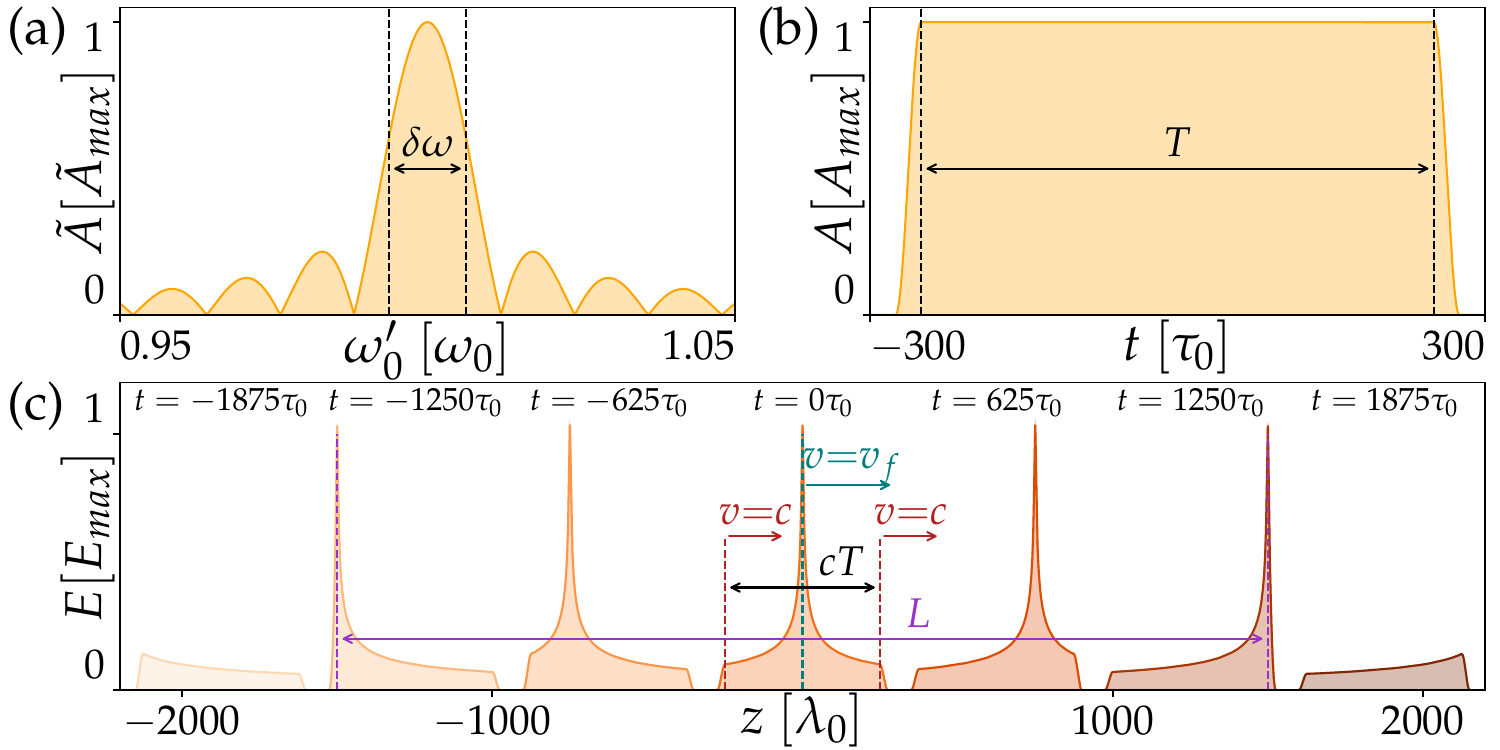}
    \caption{A finite-energy optical pulse exhibits a constant-velocity, propagation-invariant intensity peak over a finite distance. (a,b) To be truly propagation invariant, the pulse must have a flattop temporal profile. (c) The finite duration $T$ and speed $c$ of the pulse limit the distance over which the moving intensity peak is visible. The focus travels at $v_f = 1.2c$ and outruns the pulse after a distance $\smash{L = cv_f|1-\beta_f|^{-1}T}$. In this example, $T=500\tau_{0}$ where $\tau_0$ is the laser period, giving $L = 3000c
    \tau_0$. }
    \label{fig:envelope}
\end{figure}

\textbf{\textit{Physical Constraints}} -- The analysis presented above considered propagation-invariant wave packets that advect at $v_f$ forever, across all $z$ and $t$. As a result, these wave packets have infinite energy. Physically realizable optical pulses are localized in $z$ and $t$ and have finite energy. Finite-energy pulses are described by a surface that lies on the light cone in $(ck_z,ck_r,\omega)$ space. However, the condition $\Omega(\omega,k_z) = 0$ reduces a surface on the light cone to a lower-dimensional curve  [\eqref{eq:infinite-general}]. Finite-energy pulses exhibiting an arbitrary-velocity, propagation-invariant intensity peak can be constructed by superposing a continuum of wave packets with different main frequencies $\omega_0'$. The superposition assembles a surface on the light cone from the continuous set of curves $\Omega(\omega,k_z;\omega_0') \equiv \omega - \omega_{0}' - (k_z - \omega_0'/c)v_f = 0$. Symbolically, this superposition is achieved by integrating \eqref{eq:infinite-general} over $\omega_0'$:
\begin{equation}
    \begin{aligned}
    E(\mathbf{x}, t) {=} {\iint} &\tilde{A}(\omega_0') f(\mathbf{k},\omega)\ \delta[\Omega(\omega,k_z;\omega_0')]\delta(\omega-c|\mathbf{k}|)\ \\ & \times e^{i(\mathbf{k}\cdot\mathbf{x}-\omega t)}d\mathbf{k}d\omega d\omega_0'\\
    \end{aligned}
\end{equation}
where $\tilde{A}(\omega_0')$ is a spectral amplitude with a non-zero width $\delta\omega$ centered about a primary frequency $\omega_0$ [Fig.~\ref{fig:envelope}(a)]. The spectral amplitude ensures that the optical pulse has a non-zero bandwidth and finite duration $T\sim 1/\delta\omega$. The Fourier transform of $\tilde{A}(\omega_0' - \omega_0)$ determines the temporal envelope of the optical pulse $A(t)$  [Fig.~\ref{fig:envelope}(b)]. 

A propagation-invariant pulse with a finite duration $T$ and non-zero bandwidth $\delta\omega\sim1/T$ will exhibit an intensity peak moving at $v_f$ over a finite distance $L$ [Fig.~\ref{fig:envelope}(c)]. This distance, or ``focal range,'' is determined by how long it takes the moving focus to traverse the finite-duration pulse. As the relative speed between the pulse and focus is $c-v_f$, the time it takes the focus to traverse the pulse is $\smash{\tau = |1-\beta_f|^{-1}T}$. The focal range is then $\smash{L = v_f \tau = cv_f|1-\beta_f|^{-1}T}$. Figure~\ref{fig:envelope}(c) illustrates this: the temporal profile of the pulse moves at $c$, whereas the focus moves at $v_f \neq c$, implying that, after some time, the focus outruns the envelope and is no longer visible.

\textbf{\textit{Conclusions and Outlook}} -- STWPs and the IFF have the same spatiotemporal field structure despite being discovered in different contexts and synthesized through different methods. Moreover, any constant-velocity, propagation-invariant wave packet conforms to this structure---that is, the structure is universal. This universality was established by performing Lorentz boosts on generic field structures. Further analysis showed that optical pulses with finite energies can exhibit propagation-invariant intensity peaks over a finite distance, thereby verifying that propagation invariance does not require unphysical infinite-energy pulses. These results lay the foundation for a new methodology to unify independently discovered field structures. For instance, the original chromatic FF \cite{froula_spatiotemporal_2018} and axially encoded STWPs \cite{Motz21PRA} exhibit similar on-axis spectral evolution, with the former having a fixed focal velocity and the latter a fixed group velocity. A universal structure may also underlie these two realizations of space-time structured light.

The analysis presented here focused on longitudinal Lorentz boosts applied to simple solutions of Maxwell's Equations. Arbitrary Lorentz boosts applied to more elaborate solutions may reveal new structures or connections between structures of utility to applications. Furthermore, wave packets with non-constant velocities may be describable using the concepts of general relativity and non-inertial reference frames. Finally, exploring the nonlinear interaction and propagation of these pulses in different media, such as plasma, could enable a wide range of applications and fundamental measurements, further expanding the boundaries of light-matter interaction.

\textbf{Funding} -- The work of R.A. is supported by Fundação para a Ciência e Tecnologia (FCT, Portugal) grant number UI/BD/154677/2022 and by EuPRAXIA through Project 1801P.01238.1.01 EuPRAXIA PP grant number BL9/2023-IST-ID. The work of J.P.P. and D.R. is supported by the Office of Fusion Energy Sciences under Award Number DE-SC0021057. The work of A.F.A. is supported by the U.S. Office of Naval Research (ONR) under MURI award N00014-20-1-2789. 

\textbf{Disclosures} -- The authors declare no conflicts of interest.

\textbf{Data availability} -- Data available from the corresponding authors upon reasonable request.

\bibliographystyle{apsrev4-2}
%

\end{document}